\documentclass[12pt]{article}
\usepackage{epsfig}

\def\beq{\begin{equation}}
\def\eeq#1{\label{#1}\end{equation}}
\def\eeqn{\end{equation}}

\def\beqa{\begin{eqnarray}}
\def\eeqa#1{\label{#1}\end{eqnarray}}
\def\eeqan{\end{eqnarray}}

\let\bar=\overbar

\def\Dslash{\not{\hbox{\kern-4pt $D$}}}
\def\dslash{\not{\hbox{\kern-2pt $\del$}}}

\def\msb{{\bar{\ssstyle M \kern -1pt S}}}

\def\Title#1{\begin{center} {\Large {\bf #1} } \end{center}}

\begin{document}

\Title{Semileptonic $b$ to $u$ transition}

\begin{center}{\large \bf Contribution to the proceedings of HQL06,\\
Munich, October 16th-20th 2006}\end{center}

\bigskip\bigskip


\begin{raggedright}  

{\it Eunil Won\index{Won, E.}\\
Department of Physics \\
Korea University\\
136-713 Seoul, Korea}
\bigskip\bigskip
\end{raggedright}

\section{Introduction}

 The parameter $|V_{ub}|$ is one of the smallest and least known 
elements of the Cabibbo-Kobayashi-Maskawa (CKM) quark-mixing
matrix. A precise determination of $|V_{ub}|$ would significantly
improve the constraints on the unitarity triangle and provide
a stringent test of the Standard Model mechanism for
$CP$ violation. With the CKM angle $\phi_3$, $|V_{ub}|$ can constrain
the unitarity triangle from tree level processes alone.

 Experimental studies of charmless semileptonic $B$ decays can be
broadly categorized into inclusive and exclusive measurements
depending on how the final states are treated. The inclusive
method measures the decay rate $\Gamma(B \rightarrow X_u \ell \nu)$,
where $X_u$ is known as the hadronic system that does not contain
charm-quark. On the other hand, the exclusive method measures the
decay rates for exclusive final states such as $B \rightarrow \pi \ell \nu$
and $\rho \ell \nu$. Two methods give not only different efficiencies
and signal-to-background ratios, but also different theoretical 
calculations to be used in order to extract $|V_{ub}|$. Using both
approaches and comparing the results will help us verify the robustness
of the theoretical error estimation, which dominates the current uncertainty 
in the determination of $|V_{ub}|$. Progress in last few years will
be summarized in this presentation.

\section{Inclusive determination of $|V_{ub}|$}

 The inclusive semileptonic decay rates in the quark level
depend only on CKM matrix element and 
the quark mass, as shown in the following equation
\begin{eqnarray}
\Gamma(b \rightarrow u \ell \bar{\nu})
= \frac{G^2_F}{192\pi^2} |V_{ub}|^2 m^5_b
\end{eqnarray}
where $G_F$ is the Fermi constant and $m_b$ is the $b$-quark mass. The
hadronic level is easy to calculate with the framework of operator
product expansion (OPE)~\cite{ope}
and the $|V_{ub}|$ can be parametrized as 
\begin{eqnarray}
|V_{ub}| = 0.00424
\left\{
\frac{\mathcal{B}(B \rightarrow X_u \ell \nu)}{0.02}
\frac{1.61 \textrm{ps}}{\tau_b}
\right\}
^{\frac{1}{2}}
\times (1.0 \pm 0.012_\textrm{QCD} \pm 0.022_\textrm{HQE})
\end{eqnarray}
where the first error comes from the uncertainty in the calculation
of quantum chromodynamics (QCD), perturbative and non-perturbative
quantities and
the 2nd from the uncertainty in the heavy quark expansion (HQE), sensitive
to $m_b$.  This formulation~\cite{vub_parameter} has been updated in 
ICHEP06.
The main problem in the inclusive method is the background
from the $b$ to $c$ decays because the rate is approximately 50 times larger,
namely
\begin{eqnarray}
\Gamma(b \rightarrow c \ell \nu) \sim
50 \Gamma(b \rightarrow u \ell \nu).
\end{eqnarray}
One may attempt to remove this large background by applying kinematic
selection criteria but once the signal to background ratio becomes
controllable, the OPE is known to fail in such limited phase space.
In the theory side, in order to overcome this problem, various
different techniques have been developed. For example, a non-perturbative
shape function~\cite{blnp} (SF) is 
developed to extrapolate to the full phase space.
The shape function is the lightcone momentum distribution function of the
$b$-quark inside the meson. The detailed shape is not known theoretically
from the first principle and even phenominologically, the low-tail part
is least known. The shape function is needed to be determined from 
experimental data. The other approach is called dressed gluon exponentiation
(DGE) ~\cite{dge} according to the factorization properties of the 
fully differential width in inclusive decays Sudakov logarithms
exponentiate in moment space. The third approach~\cite{llr}
measures the ratio of
$|V_{ub}|/|V_{ts}|$ with the photon energy spectrum in $b \rightarrow s \gamma$
decay mode. This technique is interesting as the only residual
shape function dependence remain at the end. 

\begin{figure}[htb]
\begin{center}
\epsfig{file=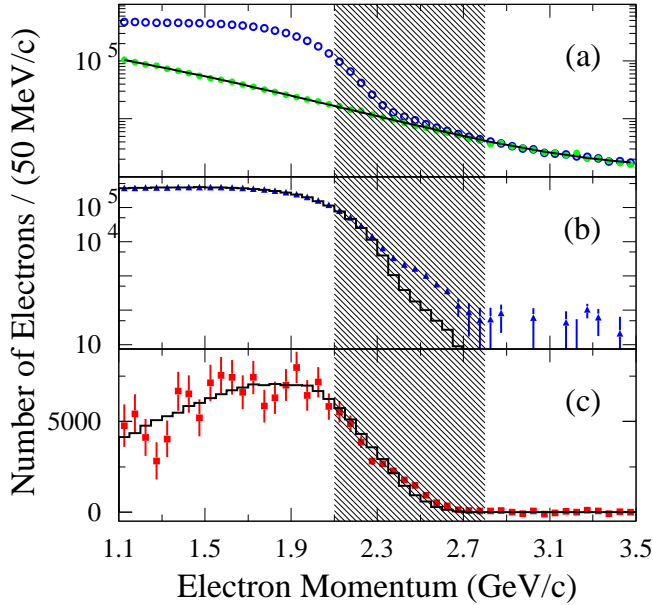,height=3.5in}
\caption{The distribution of electron momentum is 
shown~\cite{babar:lepton}. Open circles
in (a) show the on-resonance data and closed circles with a curve
in (a) show the off-resonance data where non-$B\bar{B}$ background
is included. The triangles in (b) show the data after the non
$B\bar{B}$ background subtraction and the histogram (b) show 
the simulated $B\bar{B}$ background. Closed squares and
the histogram in (c) show
the data and the simulated signal after all background subtraction,
respectively.}
\label{fig:lepton_spectrum_babar}
\end{center}
\end{figure}

\begin{itemize}
\item Lepton Endpoint Analysis
\end{itemize}
 The theoretical calculations allow for the extraction of the observed
partial $B \rightarrow X_u \ell \nu$ decay rate above a certain lepton 
momentum  to the total inclusive $B \rightarrow X_u \ell \nu$ decay rate
using the measured shape function parameters and a subsequent 
translation of the total decay rate to $|V_{ub}|$. The experimental
study was pioneered by the CLEO collaboration~\cite{cleo:lepton} and
the recent work was carried out by the BaBar 
collaboration~\cite{babar:lepton}. Fig.~\ref{fig:lepton_spectrum_babar}
(a), (b), and (c) show the electron energy spectra for various cases.
Open circles
in Fig. ~\ref{fig:lepton_spectrum_babar} (a) show 
the on-resonance data and closed circles with a curve
in Fig.~\ref{fig:lepton_spectrum_babar} (a) show 
the off-resonance data where non-$B\bar{B}$ background
is included. The triangles in Fig.~\ref{fig:lepton_spectrum_babar} 
(b) show the data after the non
$B\bar{B}$ background subtraction and the histogram in 
Fig. ~\ref{fig:lepton_spectrum_babar} (b) show 
the simulated $B\bar{B}$ background. Closed squares and
the histogram in Fig. ~\ref{fig:lepton_spectrum_babar} (c) show
the data and the simulated signal after all background subtraction,
respectively. It is obvious that the subtraction of backgrounds 
is extremely crucial in this kind of analysis.
The shaded region in Fig. ~\ref{fig:lepton_spectrum_babar} (c) 
is used for the final extraction of the $|V_{ub}|$. In the mean time,
the partial branching fraction can be obtained from the background
subtracted data and the summary from three experiment is listed in
Table~\ref{table:lepton}. Note that the Belle's result has the lowest
cut on $E_\ell$.

\begin{table}
\begin{center}
\caption{Summary of the partial branching fraction of $B \rightarrow 
X_u \ell \nu$ given the lepton energy cut.}
\begin{tabular}{rrr}
\hline
data & $\Delta \mathcal{B}(10^{-4})$ & $E_l$ (GeV) \\
\hline
CLEO   9/fb & 2.30 $\pm$ 0.15 $\pm$ 0.35 & 2.1 \\
Belle 27/fb & 8.47 $\pm$ 0.37 $\pm$ 1.53 & 1.9 \\
BaBar 80/fb & 5.72 $\pm$ 0.41 $\pm$ 0.65 & 2.0 \\
\hline
\end{tabular}
\label{table:lepton}
\end{center}
\end{table}

\begin{itemize}
\item Measurements of $m_X$, $P_+$, and $q^2$ 
\end{itemize}

In this analysis, the measurements are made with a sample of 
events where the hadronic decay mode of the tagging side $B$ meson,
$B_{tag}$, is fully reconstructed, while the semileptonic decay
of the signal side $B$ meson, $B_{sig}$, is identified by the
presence of a high momentum electron or muon. $B$ denotes both 
charged and neutral $B$ mesons. This method allows the construction
of the invariant masses of the hadronic ($M_X$) and leptonic
($\sqrt{q^2}$) system in the semileptonic decay, and the variable
$P_+=E_X - |\vec{p}_X|$ where $E_X$ is the energy and $|\vec{p}_X|$
the magnitude of the three-momentum of the hadronic system.
These inclusive kinematic variables can be used to separate the
$B\rightarrow X_u \ell \nu$ decays from the much more abundant 
$B\rightarrow X_c \ell \nu$ decays. Three competing kinematic
regions were proposed by theoretical studies~\cite{blnp,bll}, based
on the three kinematic variables, and are directly compared by
this analysis. 

\begin{figure}[htb]
\begin{center}
\epsfig{file=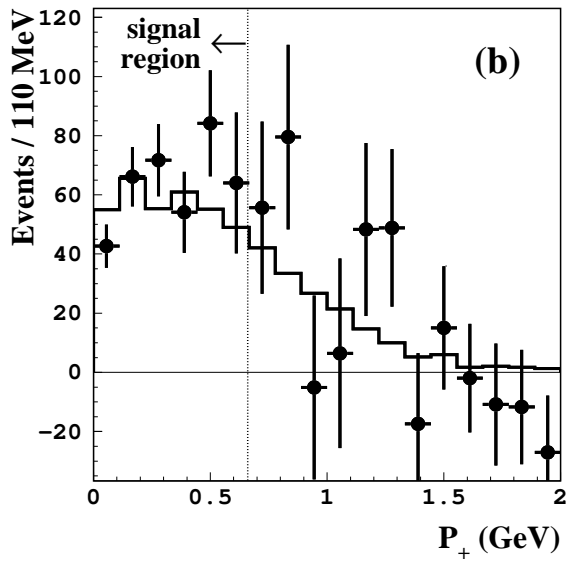,height=3.0in}
\caption{
The $P_+$ distribution (symbols with error bars) 
after subtracting $B \rightarrow X_c \ell \nu$ background,
with fitted $B \rightarrow X_u \ell \nu$ contribution (histogram). 
}
\label{fig:belle_p+}
\end{center}
\end{figure}

The value of $|V_{ub}|$ is extracted using recent
theoretical calculation~\cite{blnp,bll} that include all the
current known contributions. Results from the Belle 
experiment using a data set of 253/fb~\cite{belle:p+}
is the first one to use $P_+$. Figure~\ref{fig:belle_p+} show the
distribution of the variable $P_+$. Signal region is defined by
$P_X < 0.66$ GeV and shows a clear indication of enhancement of 
the signal in Fig.~\ref{fig:belle_p+}. BaBar did also similar
analysis but only with $M_X$ and $q^2$~\cite{babar:m_x}. The measured
partial branching fractions from both experiments for different
phase space values are summarized in Table~\ref{table:p+}. Note
that the errors are larger than those appeared in the endpoint 
analyses results.

\begin{table}
\begin{center}
\caption{Summary of the partial branching fraction of $B \rightarrow 
X_u \ell \nu$ given the lepton energy cut.}
\begin{tabular}{rcr}
\hline
data & Phase Space & $\Delta \mathcal{B}$ (10$^{-4}$) \\
\hline
             & $M_X < $ 1.7             & 12.4 $\pm$ 1.1 $\pm$ 1.0 \\
Belle 253/fb & $M_X < $ 1.7, $q^2 >$ 8  &  8.4 $\pm$ 0.8 $\pm$ 1.0 \\
             & $P_+ <$ 0.66             & 11.0 $\pm$ 1.0 $\pm$ 1.6 \\
\hline
BaBar 211/fb & $M_X < $ 1.7, $q^2 >$ 8  &  8.7 $\pm$ 0.9 $\pm$ 0.9 \\
             &              & (preliminary)  \\
\hline
\end{tabular}
\label{table:p+}
\end{center}
\end{table}

\begin{itemize}
\item Extraction of $|V_{ub}|$
\end{itemize}

 In order to extract the $|V_{ub}|$ from the partial branching fraction
measurements described so far, one has to know the distribution of the
shape function. The photon energy spectrum in $B \rightarrow X_s \gamma$
provides access to such distribution function of the $b$ quark inside
the $B$ meson~\cite{neubert:bquark}. The knowledge of this shape function
is a crucial input to the extraction of $|V_{ub}|$  from inclusive
semileptonic $B \rightarrow X_u \ell \nu$ measurements. Both Belle and
Babar fit the spectrum to theoretical predictions in order to extract
the $|V_{ub}|$.
 
\begin{figure}[htb]
\begin{center}
\epsfig{file=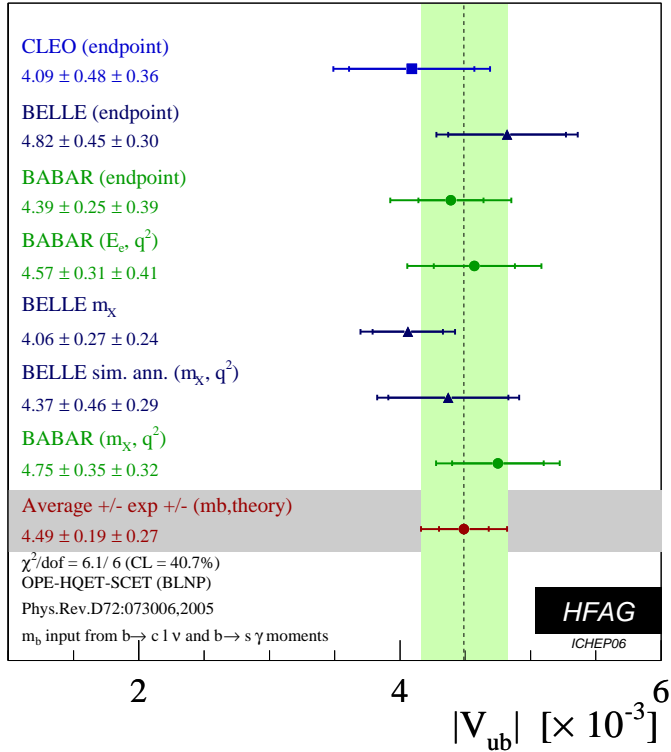,height=4.0in}
\caption{
Summary of the measurements of $|V_{ub}|$ within BLNP
framework from various experiments are shown. The averaged value is
$|V_{ub}|_{\textrm{BLNP}}$ = ($4.49 \pm$ $0.19_{\textrm{exp}} \pm$
$0.27_{\textrm{theory}}$) $\times 10^{-3}$.
}
\label{fig:vub_allMoments}
\end{center}
\end{figure}

 Using the heavy quark parameter values $m_b$(SF) = 4.60 $\pm$ 0.04 GeV and
$\mu^2_\pi$(SF) = 0.20 $\pm$ 0.04 GeV$^2$, the value of $|V_{ub}|$ from 
various experimental results is
extracted within ``BLNP'' framework~\cite{blnp}. This is 
done by the heavy flavour averaging group~\cite{hfag} (HFAG) and the result is
\begin{eqnarray}
|V_{ub}|_{\textrm{BLNP}} = (4.49 \pm 0.19_{\textrm{exp}} \pm
0.27_{\textrm{theory}}) \times 10^{-3}.
\end{eqnarray}
Note that the total error in percentage is 
\begin{eqnarray}
\delta |V_{ub}|_{\textrm{BLNP}} = \pm 7.3 \% 
\end{eqnarray}
and the result indicates that the precision of the
measurement is still limited by theoretical uncertainty but the
value of the 
error is not far from the experimental error. The individual result
from experiments is shown in Fig.~\ref{fig:vub_allMoments}.
Note that the best measurement is from the endpoint analysis at present.
Contributions to
the experimental and theoretical errors can be found in Table~\ref{table:blnp}.
The contribution from sub-leading SF, for example, is known to be hard
to reduce from the present value.
\begin{table}
\begin{center}
\caption{Summary of the contributions to the experimental and 
theoretical errors of $|V_{ub}|$ in the BLNP framework.}
\begin{tabular}{cc}
\hline
Source & contribution (\%) \\ 
\hline
statistical      & 2.2 \\
Expt. systematic & 2.8 \\
$b \rightarrow c\ell \nu$ model & 1.9 \\
$b \rightarrow u\ell \nu$ model & 1.6 \\
HQ parameters & 4.2 \\
sub-leading SF & 3.8 \\
Weak Annihilation & 1.9 \\
\hline
\end{tabular}
\label{table:blnp}
\end{center}
\end{table}

 One can also extract $|V_{ub}|$ using the DGE framework~\cite{dge}. 
This is also done by HFAG and the result is
\begin{eqnarray}
|V_{ub}|_{\textrm{DGE}} = (4.46 \pm 0.20_{\textrm{exp}} \pm
0.20_{\textrm{theory}}) \times 10^{-3}.
\end{eqnarray}
with the parameter $m_b$(MS) = 4.20 $\pm$ 0.04 GeV. Note that
the result is in good agreement with the result with BLNP method, and
is remarkable as 
two theoretical methods use rather different approach in their
calculations. The contribution of the error is also listed in
Table~\ref{table:dge}.
\begin{table}
\begin{center}
\caption{Summary of the contributions to the experimental and 
theoretical errors of $|V_{ub}|$ in the DGE framework.}
\begin{tabular}{cc}
\hline
Source & contribution (\%) \\ 
\hline
statistical      & 1.8 \\
Expt. systematic & 2.5 \\
$b \rightarrow c\ell \nu$ model & 2.3 \\
$b \rightarrow u\ell \nu$ model & 2.3 \\
$m_b$ (R\_CUT) & 1.2 \\
$\alpha_s$ (R\_CUT) & 1.0 \\
Total semiloptonic width & 3.0 \\
DGE theory & 2.9 \\
\hline
\end{tabular}
\label{table:dge}
\end{center}
\end{table}
The sharing of the error is also similar to the case of the BLNP framework.
One thing to note is that the weak annihilation is not taken into account
in this case.

\begin{figure}[htb]
\begin{center}
\epsfig{file=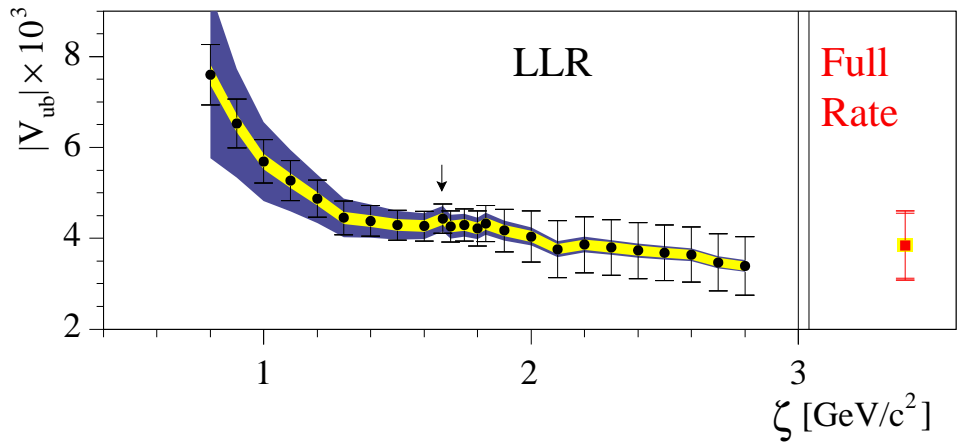,height=2.3in}
\caption{
$|V_{ub}|$ as a function of $\zeta$ with the LLR method (left) and
for the determination with the full rate measurement (right). The
error bars indicate the statistical uncertainty. They are correlated
between the points and get larger for larger $\zeta$ due to larger
background from $B \rightarrow X_c \ell \nu$. The total shaded are 
illustrates the theoretical uncertainty; the inner light shaded area
indicates the perturbative share of the uncertainty. 
The arrow indicates $\zeta$ = 1.67 GeV/c$^2$.
}
\label{fig:babar_llr_vub}
\end{center}
\end{figure}

 On the other hand, BaBar explored alternative methods in obtaining
$|V_{ub}|$. Leibovich, Low, and Rothstein (LLR)~\cite{llr} 
have presented a prescription
to extract $|V_{ub}|$ with reduced model dependence from either the
lepton energy or the hadronic mass $m_X$~\cite{babar:llr}. 
The calculations of LLR
are accurate up to corrections of order $\alpha^2_s$ and 
$(\Lambda m_B/(\zeta m_b))^2$, where $\zeta$ is the experimental maximum
hadronic mass up to which the $B \rightarrow X_u \ell \nu$ decay rate
is determined and $\Lambda \sim \Lambda_{\textrm{QCD}}$. This method
combines the hadronic mass spectrum, integrated below $\zeta$, with
the high-energy end of the measured differential
$B \rightarrow X_s \gamma$ photon energy spectrum via the calculations
of LLR. The measured $|V_{ub}|$ as a function of $\zeta$ is shown in
Fig.~\ref{fig:babar_llr_vub}. The small arrow indicates the value of 
$\zeta$ that is used for the cut, $\zeta$ = 0.67 GeV/c$^2$. At this point,
the measured value becomes
\begin{eqnarray}
|V_{ub}| = (4.43 \pm 0.38_{\textrm{stat}} \pm
0.25_{\textrm{syst}} \pm 0.29_{\textrm{th}}) \times 10^{-3}.
\end{eqnarray}

Another approach that BaBar chooses to reduce the model dependence 
is to measure the  $B \rightarrow X_u \ell \nu$ rate over the entire $M_X$
spectrum. Since no extrapolation is necessary to obtain the full rate, 
systematic uncertainties from $m_b$ and Fermi motion are much reduced.
Perturbative corrections are known to order $\alpha_s$. The rate of
$B \rightarrow X_u \ell \nu$ is extracted from the hadronic mass
spectrum up to $\zeta$ = 2.5 GeV/c$^2$ which corresponds to about 96 \%
of the simulated hadronic mass spectrum, and find $|V_{ub}|$ = (3.84 $\pm$ 
0.70$_{\textrm{stat}}$ $\pm$
0.30$_{\textrm{syst}}$ $\pm$
0.10$_{\textrm{th}} )$ $\times$ 10$^{-3}$, using the average $B$ lifetime of
$\tau_B$ = (1.604 $\pm$ 0.012) ps. The current uncertainties on the 
$B \rightarrow X_s \gamma$ photon energy spectrum limit the
sensitivity with which the behavior at high $\zeta$ can be probed. These two
new results are consistent with previous measurements but have substantially 
smaller uncertainties from $m_b$ and the modeling for Fermi motion of
the $b$ quark inside the $B$ meson. Both techniques are based on theoretical
calculations that are distinct from other calculations normally 
employed to extract $|V_{ub}|$ and, thus, provide a complementary 
determination of $|V_{ub}|$.

\section {Exclusive determination of $|V_{ub}|$}

 In this section, we discuss the extraction of $|V_{ub}|$ from 
exclusive decays such as $B \rightarrow (\pi,\rho,\omega) \ell \nu$.
For $B^0 \rightarrow \pi^- \ell^+ \nu$ decays, the differential decay
rate becomes
\begin{eqnarray}
\frac{d\Gamma(B \rightarrow \pi \ell \nu)}{dq^2}
= \frac{G^2_F}{24\pi^3}|V_{ub}|^2p^3_\pi|f_+(q^2)|^2
\end{eqnarray}
where $f_+(q^2)$ is a form factor and $q^2$ is the squared invariant mass
of the $\ell^+ \nu$ system. Only shape of $f_+(q^2)$ can be measured
experimentally. Its normalization is provided by theoretical 
calculations which currently suffer from relatively large uncertainties
and, often, do not agree with each other. As a result, the normalization 
of the $f_+(q^2)$ form factor is the largest source of uncertainly in
the extraction of $|V_{ub}|$ from the $B^0 \rightarrow \pi^- \ell^+ \nu$
branching fraction. Values of $f_+(q^2)$ for
$B^0 \rightarrow \pi^- \ell^+ \nu$
decays are provided by unquenched~\cite{lattice1,lattice2} and 
quenched~\cite{quenched} lattice QCD calculations, presently reliable
only at high $q^2$ ($>$ 16 GeV$^2$/c$^4$), and by Light Cone Sum
Rules calculations~\cite{lcsr} (LCSR) based on approximations only valid
at low $q^2$ ($<$ 16 GeV$^2$/c$^4$), as well as by a quark model~\cite{qmodel}. 
The QCD theoretical predictions are at present more precise for
$B^0 \rightarrow \pi^- \ell^+ \nu$ decays than for other
exclusive $B \rightarrow X_u \ell \nu$ decays. Experimental data can
be used to discriminate between the various calculations by measuring
the $f_+(q^2)$ shape precisely, thereby leading to a smaller 
theoretical uncertainty on $|V_{ub}|$.

\begin{figure}[htb]
\begin{center}
\epsfig{file=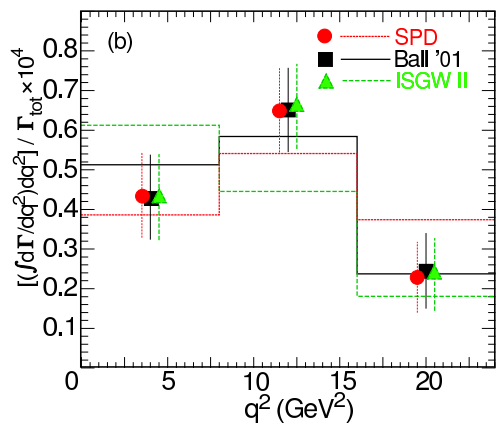,height=2.3in}
\caption{
Fit to $d\Gamma/dq^2$ on exclusive $B\rightarrow \pi^- \ell^+ \nu$
decays.
}
\label{fig:cleo_pilnu_dq2}
\end{center}
\end{figure}

 For the recoiling the other side of the $B$ meson, we may require
nothing (untagged) or require indication of semileptonic decay (D$^{(*)}
\ell \nu$ tag), and require hadronic decay (full reconstruction tag). 
We review recent progress on experimental studies with different tagging
methods listed above.

\begin{itemize}
\item Untagged $B \rightarrow \pi \ell \nu$
\end{itemize}

CLEO pioneered 
the measurement of $V_{ub}$ with exclusive decays~\cite{cleo:untagged}. They
perform a simultaneous maximum likelihood fit in $\Delta E$ and $M_{m\ell \nu}$
to seven sub-modes: $\pi^\pm$, $\pi^0$, $\rho^\pm$, 
$\omega/\eta \rightarrow pi^+\pi^-\pi^0$ , and
$\eta \rightarrow \gamma \gamma$. In the fit they used isospin symmetry to
constrain the semileptonic widths
$\Gamma^{\textrm{SL}}(\pi^\pm)$ = $2\Gamma^{\textrm{SL}}$ ($\pi^0$) and
$\Gamma^{\textrm{SL}}(\rho^\pm)$ = $2\Gamma^{\textrm{SL}}$ ($\rho^0$) 
$\sim$ 
$2\Gamma^{\textrm{SL}}$ ($\omega$), where the final approximate equality
is inspired by constituent quark symmetry. Signals for $\pi$ and $\rho$
are extracted separately in three $q^2$ bins. Given form factors from
theory, they extracted $|V_{ub}|$ from a fit to $d\Gamma/dq^2$, and it
is shown in Fig.~\ref{fig:cleo_pilnu_dq2}. From their fit, 
we see that ISGW2~\cite{qmodel}  is least favored. Combining 
$B \rightarrow \pi \ell \nu$ and
$B \rightarrow \rho \ell \nu$ results, CLEO found that

\begin{eqnarray}
|V_{ub}| = (3.17 \pm 0.17 (\textrm{stat}) ^{+0.16}_{-0.17}  (\textrm{syst})
^{+0.53}_{-0.39} (\textrm{theo}) \pm 0.03_{\textrm{FF}} ) \times 10^{-3}.
\end{eqnarray}
\begin{figure}[htb]
\begin{center}
\epsfig{file=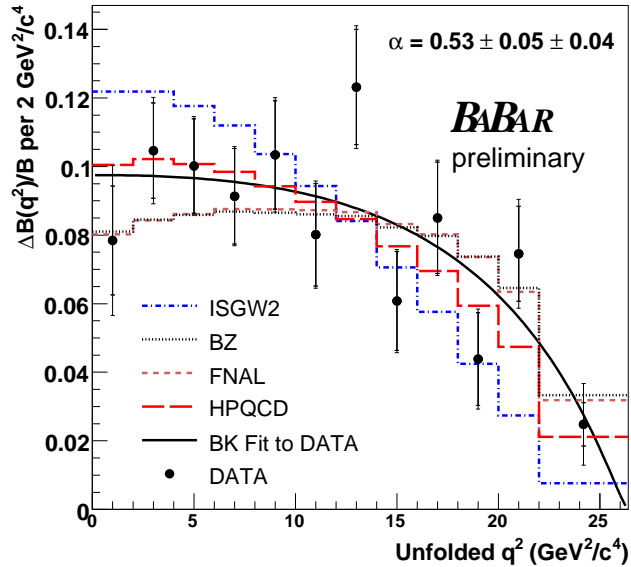,height=3.3in}
\caption{
Differential decay rate formula fitted to the normalized partial
$\Delta \mathcal{B}/\mathcal{B}$ spectrum in 12 bins of $q^2$.
The smaller error bars are statistical only while the larger error
bars include statistical and systematic uncertainties. The BK 
parametrization (solid black curve) reproduces the data quite well
($\chi^2$=8.8 for 11 degrees of freedom) with the parameter 
$\alpha$ = 0.53 $\pm$ 0.05 $\pm$ 0.04. The data are also compared
to LCSR calculations (dotted line), unquenched LQCD calculations
(long dashed line and short dashed line) and the ISGW2 quark model 
(dash-dot line).
}
\label{fig:babar_FplusFit}
\end{center}
\end{figure}

 BaBar also did similar but improved analysis~\cite{babar:untagged}. 
More accurate value of $q^2$
was obtained in the so-called $Y$-average frame where the
pseudo-particle $Y$ has a four-momentum 
defined by $P_Y \equiv (P_\pi + P_\ell)$.
The angle $\theta_{YT}$ between the directions of the $p^*_B$ and $p^*_Y$
momenta in the $\Upsilon(4S)$ rest frame can be determined assuming
energy-momentum conservation in a semileptonic $B \rightarrow Y \nu$
decay. The use of the $Y$-averaged frame yields a $q^2$ resolution that
is approximately 20 \% bettern than what is obatined in the usual
$\Upsilon(4S)$ frame where the $B$ meson is assumed to be at rest. 
They also fit the 
$\Delta \mathcal{B}/\mathcal{B}$ spectrum using a probability
density function based on
the $f_+(q^2,\alpha)$ parametrization of Becirevic-Kaidalov~\cite{bk} (BK).
The normalized 
$\Delta \mathcal{B}/\mathcal{B}$ distribution is shown 
in Fig.~\ref{fig:babar_FplusFit}, together with the result of a
$f_+(q^2)$ shape fit using the BK parametrization and theoretical prediction.
The obtain a value of $\alpha$ = 0.53 $\pm$ 0.05 $\pm$ 0.04. BaBar data
are clearly incompatible with the ISGW2 quark model, which is in agreement
with what CLEO data indicated before. The extraction of $|V_{ub}|$
is carried out from the partial branching fractions using
$|V_{ub}|$ = $\sqrt{\Delta\mathcal{B}/(\tau^0_B\Delta \zeta)}$, where
$\tau^0_B$ = (1.536 $\pm$ 0.014) ps is the $B^0$ lifetime and 
$\Delta \zeta$ is the normalized partial decay
rate predicted by various form factor calculations. For the LCSR calculations 
with $q^2$ $<$ 16 GeV$^2$/c$^4$, $|V_{ub}|$ = (3.6 $\pm$ 0.1 $\pm$ 0.1
$^{+0.6}_{-0.4})$ $\times$ 10$^{-3}$ is obtained. For the HPQCD and 
and FNAL lattice calculations with $q^2$ $>$ 16 GeV$^2$/c$^4$, 
$|V_{ub}|$ = (4.1 $\pm$ 0.2 $\pm$ 0.2
$^{+0.6}_{-0.4})$ $\times$ 10$^{-3}$ 
and
$|V_{ub}|$ = (3.6 $\pm$ 0.2 $\pm$ 0.2
$^{+0.6}_{-0.4})$ $\times$ 10$^{-3}$ 
are obtained, respectively. Note that all are in good agreement within
given uncertainty and this gives us confidence in exclusive measurement
of $|V_{ub}|$. 

\begin{itemize}
\item $D^{(*)} \ell \nu$ tag
\end{itemize}

\begin{figure}[htb]
\begin{center}
\epsfig{file=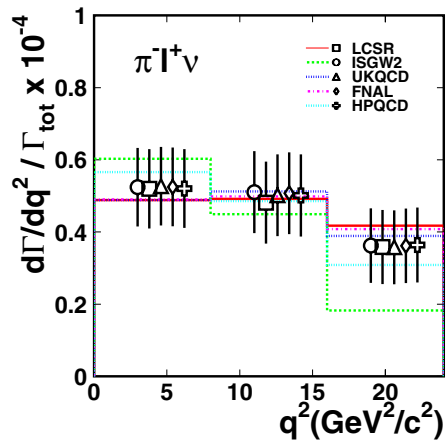,height=2.3in}
\caption{
Extracted $q^2$ distribution for $B^0 \rightarrow \pi^- \ell^+ \nu$
decays. Data points are shown for different form factor models used
to estimate the detection efficiency. Lines are for the best fit of the
form factor shapes to the obtained $q^2$ distribution.
}
\label{fig:belle_q2_dtag}
\end{center}
\end{figure}

 Belle presented measurements of $B^0 \rightarrow \pi^-/\rho^- \ell^+ \nu$
and $B^+ \rightarrow \pi^0/\rho^0 \ell^+ \nu$ decays using 
$B \rightarrow D^{(*)} \ell \nu$ tagging~\cite{belle_dtag}. 
They reconstruct the 
entire decay chain from the $\Upsilon(4S)$ $\rightarrow$ 
$B_{\textrm{sig}} B_{\textrm{tag}}$, $B_{\textrm{sig}}$ $\rightarrow$
$\pi/\rho \ell \nu$ and $B_{\textrm{tag}} \rightarrow$ 
$D^{(*)} \ell \nu$ tag with several $D^{(*)}$ sub-modes.
The back-to-back correlation of the two $B$ mesons in the $\Upsilon(4S)$
rest frame allows us to constrain the kinematics of the double semileptonic 
decay. The signal is reconstructed in four modes, 
$B^0 \rightarrow \pi^-/\rho^- \ell^+ \nu$ and 
$B^+ \rightarrow \pi^0/\rho^0 \ell^+ \nu$. Yields and branching fractions
are extracted from a simultaneous fit of the $B^0$ and $B^+$ samples
in three intervals of $q^2$, accounting for cross-feed between modes
as well as other backgrounds. Belle applied this methods to 
$B \rightarrow \pi/\rho \ell \nu$ decays for the first time,
and have succeeded in reconstructing these decays with significantly
improved signal-to-noise ratios compared to the $\nu$-reconstruction 
method.  With the data of 253 fb$^{-1}$, Belle extracted branching
fractions and $|V_{ub}|$. Figure~\ref{fig:belle_q2_dtag} shows the
$q^2$ distribution for the decay
$B^0 \rightarrow \pi^-/\rho^- \ell^+ \nu$. The error bars are too
large to reject any of form factor models for the moment. 
Table~\ref{table:belle_q2} summarized 
the results with two different lattice calculations. This gives
about 13 \% experimental uncertainty on $|V_{ub}|$, currently 
dominated by the statistical error of 11 \%. By accumulating more
integrated luminosity, a measurement with errors below 10 \% is
feasible. With improvements to unquenched LQCD calculations,
the present method may provide a precise determination of $|V_{ub}|$. 
\begin{table}
\begin{center}
\caption{Summary of $|V_{ub}|$ measurements by the Belle collaboration
with different lattice calculations.}
\begin{tabular}{ccc}
\hline
& $q^2$ GeV$^2$/c$^4$ & $|V_{ub}|$ $\times$ 10$^{-3}$  \\ 
\hline
FNAL  & $>$ 16 & 3.60 $\pm$ 0.41 $\pm$ 0.20 $\pm$ $^{0.62}_{-0.41}$ \\
HPQCD & $>$ 16 & 4.03 $\pm$ 0.46 $\pm$ 0.22 $\pm$ $^{0.59}_{-0.41}$ \\
\hline
\end{tabular}
\label{table:belle_q2}
\end{center}
\end{table}

 The BaBar collaboration also did similar analysis~\cite{babar_dtag} 
based on 211 fb$^{-1}$ of data. They also included the hadronic decay
of $B$ mesons for tagging. The detailed techniques are similar 
to what to be explained
later. The summary of the measurement of $|V_{ub}|$ can be found
in Table~\ref{table:babar_sltag}. All the results are in good agreement
with the results from the Belle experiment.
\begin{table}
\begin{center}
\caption{Summary of $|V_{ub}|$ measurements by the BaBar collaboration
with different theoretical calculations.}
\begin{tabular}{ccc}
\hline
& $q^2$ GeV$^2$/c$^4$ & $|V_{ub}|$ $\times$ 10$^{-3}$  \\ 
\hline
Ball-Zwicky & $<$ 16 & 3.2 $\pm$ 0.2 $\pm$ 0.1 $^{+0.5}_{-0.4}$ \\
HPQCD & $>$ 16 & 4.5 $\pm$ 0.5 $\pm$ 0.3 $^{+0.7}_{-0.5}$ \\
FNAL  & $>$ 16 & 4.0 $\pm$ 0.5 $\pm$ 0.3 $^{+0.7}_{-0.5}$ \\
APE~\cite{ape}   & $>$ 16 & 4.1 $\pm$ 0.5 $\pm$ 0.3 $^{+1.6}_{-0.7}$ \\
\hline
\end{tabular}
\label{table:babar_sltag}
\end{center}
\end{table}

\begin{itemize}
\item Full reconstruction tag
\end{itemize}

\begin{figure}[htb]
\begin{center}
  \mbox{
    \epsfig{file=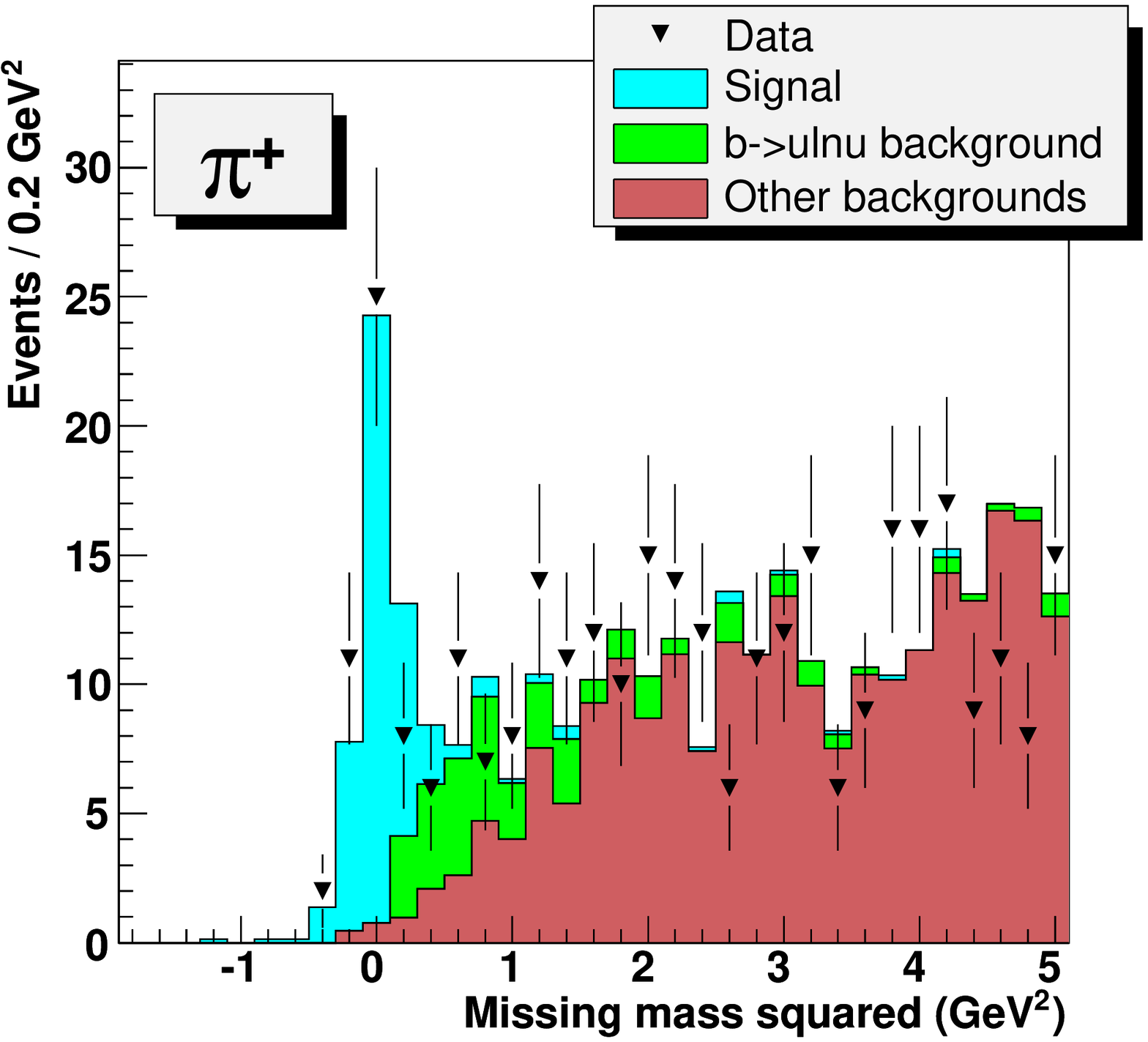,height=2.3in} \quad
    \epsfig{file=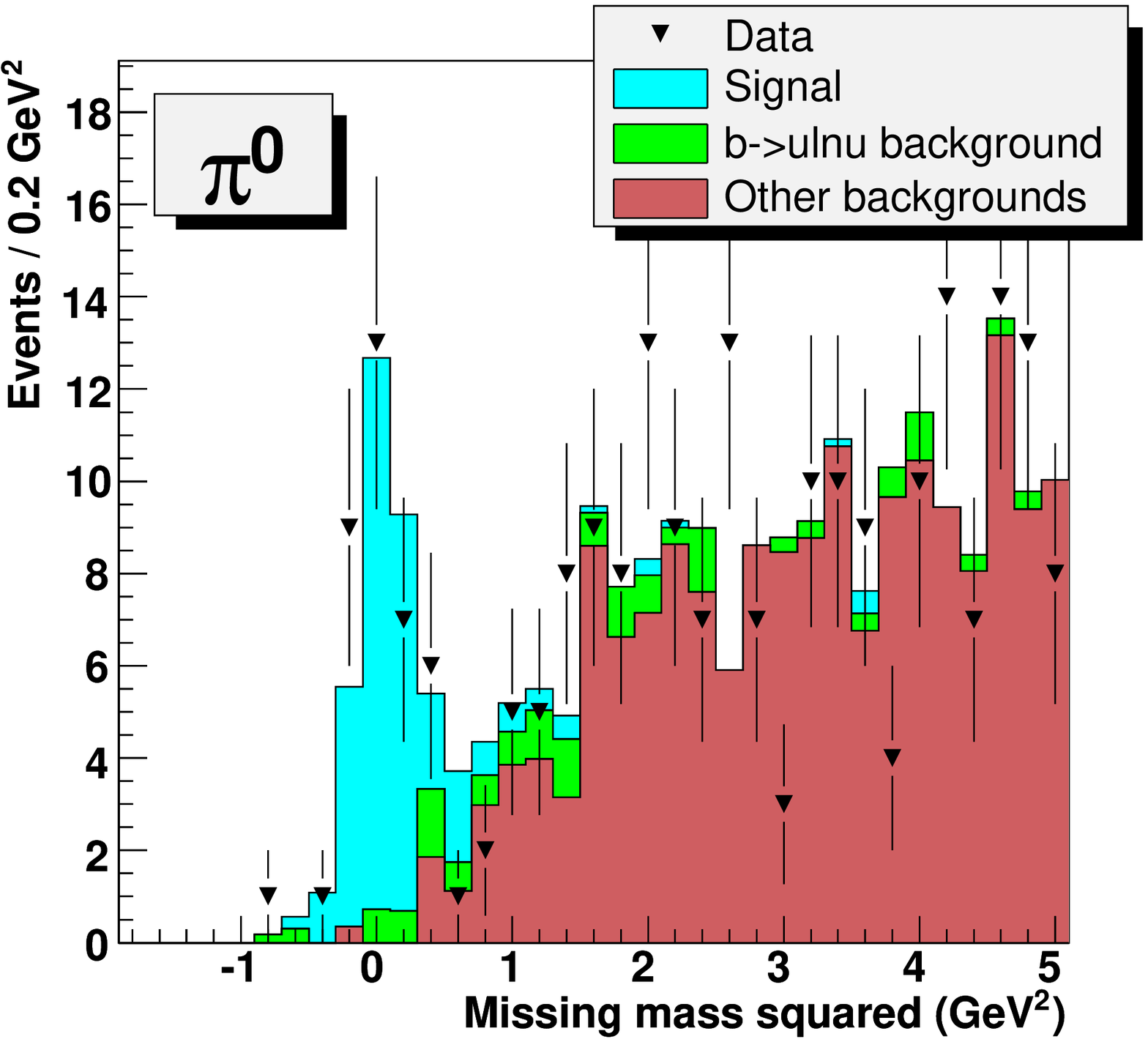,height=2.3in} 
  }
\caption{
Missing mass squared ($M^2_{\textrm{miss}}$) distributions after all cuts,
for $B \rightarrow \pi^+ \ell \nu$ on the left and
$B \rightarrow \pi^0 \ell \nu$ on the right.
Data is indicated by the points with error bars. The blue histogram 
(lightest shade in greyscale) show the fitted prediction based on
the LCSR model. The green histogram (middle shade in greyscale) shows the 
fitted $b \rightarrow u\ell \nu$ background contribution. The crimson
histogram (darkest shade in greyscale) shows the fitted background
contribution from other sources.
}
\label{fig:belle_fulltag}
\end{center}
\end{figure}

 In this method~\cite{belle:fulltag}, 
Belle fully reconstructs one of the two $B$ mesons from
$\Upsilon(4S)$ decay ($B_{\textrm{tag}}$) in one of the following hadronic
decay modes, 
$B^-$ $\rightarrow$ $D^{(*)0} \pi^-$,
$B^-$ $\rightarrow$ $D^{(*)0} \rho^-$,
$B^-$ $\rightarrow$ $D^{(*)0} a_1^-$,
$B^-$ $\rightarrow$ $D^{(*)0} D^{(*)-}_s$,
$B^0$ $\rightarrow$ $D^{(*)+} \pi^-$,
$B^0$ $\rightarrow$ $D^{(*)+} \rho^-$,
$B^0$ $\rightarrow$ $D^{(*)+} a_1^-$,
or
$B^0$ $\rightarrow$ $D^{(*)+} D^{(*)-}_s$.
As was done in other $B$ meson analyses,
decays are identified on the basis of the proximity of the beam-energy
constrained mass $M_{\textrm{bc}}$ and $\Delta E$ to their nominal
values of the $B$ meson rest mass and zero, respectively.
If multiple tag candidates are found, the one with values of $M_{\textrm{bc}}$
and $\Delta E$ closest to nominal is chosen. Events with a $B_{\textrm{tag}}$
satisfying the selections $M_{\textrm{bc}}$ $>$ 5.27 GeV/c$^2$ and 
$-0.08$ $<$ $\Delta E$ $<$ 0.06 GeV are retained. The charge of the 
$B_{\textrm{tag}}$ candidate is necessarily restricted to
$Q_{\textrm{tag}}$ = 0 
or 
$Q_{\textrm{tag}}$ $\pm$ 1 by demanding that it is consistent with
one of the above decay modes. Reconstructed charged tracks and 
electromagnetic clusters which are not associated with the
$B_{\textrm{tag}}$ candidate are used to search for the signal
$B$ meson decays of interest recoiling against the $B_{\textrm{tag}}$.
After all cuts to enhance the signal are applied,
remaining data are projected as
a function of missing mass squared ($M_{\textrm{miss}}$), as shown
in Fig.~\ref{fig:belle_fulltag}. As is demonstrated in the figure,
a high purity signal can be obtained with this method. A preliminary
branching fractions are obtained as
\begin{eqnarray}
\mathcal{B} (B \rightarrow \pi^- \ell \nu) &=&
1.49 \pm 0.26 \textrm{(stat)} \pm 0.06 \textrm{(syst)} \times 10^{-4},
\\
\mathcal{B} (B \rightarrow \pi^0 \ell \nu) &=&
0.86 \pm 0.17 \textrm{(stat)} \pm 0.06 \textrm{(syst)} \times 10^{-4}.
\end{eqnarray}
Whilst the statistical precision of these measurements is limited at present,
the potential power of the full reconstruction tagging method, when it can be
used with larger accumulated $B$-factory data samples in the future,
can clearly be seen.


\section{Conclusions}

 We have discussed recent progress in measurements of $|V_{ub}|$. At
present, the total error from the inclusive measurement is
approximately 7 \%. The theoretical uncertainty in it is still larger than
the experimental uncertainty, but not by a lot anymore. The shape function,
describing the Fermi motion of the $b$-quark inside the meson still
remains a big issue in extracting $|V_{ub}|$, but different approaches
produce consistent numerical values indicating the problem is 
well understood. Also, BaBar's new approach based on  
Leibovich,  Low and Rothstein~\cite{llr} may look promising in future as it
has residual SF dependence only.

 The exclusive measurements of $|V_{ub}|$ gives the total error to be
greater than 10 \% at present and the form factor is the main issue in this
field.  The untagged analyses are the most precise ones at present.
The theoretical uncertainty is still dominant in exclusive measurements.

\bigskip
I am grateful to
Masahiro Morii from the BaBar experiment who kindly provided
complete information on their experimental results.


\begin{thebibliography}{99}


\bibitem{ope}
J. Chay, H. Georgi, and B. Grinstein, Phys. Lett. B {\bf 247}, 399 (1990);
M. A. Shifman and M. B. Voloshin, Sov. J. Nucl. Phys. {\bf 41}, 120 (1985);
I. I. Bigi {\it et al.}, Phys. Lett. B {\bf 293}, 430 (1992);
{\bf 297}, 430(E) (1992);
Phys. Rev. Lett. {\bf 71}, 496 (1993);
A. V. Manohar and M. B. Wise, Phys. Rev. D {\bf 49}, 1310 (1994).

\bibitem{vub_parameter} N. Uraltsev, Int. J. Mod. Phys. A {\bf 14},
4641 (1999); A. H. Hoang, Z. Ligeti, and A. V. Manohar, Phys. Rev.
D {\bf 59}, 074017 (1999); T. van Ritbergen, Phys. Lett. B {\bf 454},
353 (1999).

\bibitem{blnp}
B. O. Lange, M. Neubert and G. Paz, Phys. Rev. D {\bf 72}, 073006 (2005);


\bibitem{dge}
J. R. Anderson and E. Gardi, J. High Energy Phys. {\bf 0601}, 097 (2006).

\bibitem{llr}
A. K. Leibovich, I. Low and I. Z. Rothstein, Phys. Lett. B {\bf 486}, 86 (2000).


\bibitem{cleo:lepton}
The CLEO collaboration, Phys. Rev. Lett. {\bf 88}, 231803 (2002).

\bibitem{babar:lepton}
The BaBar collaboration, Phys. Rev. D {\bf 73}, 012006 (2006).

\bibitem{bll}
C. W. Bauer, Z. Ligeti and M. E. Luke, Phys. Rev. D {\bf 64}, 113004 (2001).

\bibitem{belle:p+}
The Belle collaboration, Phys. Rev. Lett. {\bf 95}, 241801 (2005).

\bibitem{babar:m_x}
The BaBar collaboration, hep-ex/0507017.

\bibitem{neubert:bquark}
M. Neubert, Eur. Phys. J. C {\bf 40}, 165 (2005); Phys. Lett. B {\bf 612},
13 (2005); S. W. Bosch, M. Neubert, and G. Paz, J. High Energy Phys.
 {\bf 11}, 073  (2004).

\bibitem{hfag}
htp://www.slac.stanford.edu/xorg/hfag.

\bibitem{babar:llr}
The BaBar collaboration, Phys. Rev. Lett. {\bf 96}, 222801 (2006).

\bibitem{lattice1}
The HPQCD collaboration, E. Gulez {\it et al.}, Phys. Rev. {\bf D73}, 074502 (2006).

\bibitem{lattice2}
The FNAL collaboration, M. Okamoto {\it et al.}, Nucl. Phys. Proc. Suppl. 
{\bf 140} 461 (2005).

\bibitem{quenched}
A. Abada {\it et al.}, Nucl. Phys. {\bf D619}, 565 (2001).
 
\bibitem{lcsr}
P. Ball, R. Zwicky, Phys. Rev. {\bf D71}, 014015 (2005).

\bibitem{qmodel}
D. Scora, N. Isgur, Phys. Rev. {\bf D52}, 2783 (1995).

\bibitem{cleo:untagged}
The CLEO collaboration, Phys. Rev. {\bf D68}, 072003 (2003).

\bibitem{babar:untagged}
The BaBar collaboration, hep-ex/0607060.

\bibitem{bk}
D. Becirevic and A. B. Kaidalov, Phys. Lett. {\bf B478}, 417 (2000).

\bibitem{belle_dtag}
The Belle collaboration, hep-ex/0604024.

\bibitem{babar_dtag}
The BaBar collaboration, hep-ex/0607089.

\bibitem{ape}
A. Abada {\it et al.}, Nucl. Phys. {\bf B619}, 565 (2001).

\bibitem{belle:fulltag}
The Belle collaboration, hep-ex/0610054. 

\end{thebibliography}
\end{document}